\documentclass[apjl]{emulateapjDCE}

\usepackage{ulem} % AV
\usepackage{color} % AV

\newcommand{\cutoff}{\alpha_\mathrm{cut}}
\newcommand{\Facc}{F_\mathrm{acc}}

\newcommand{\Iic}{I_\mathrm{IC}}
\newcommand{\Isyn}{I_\mathrm{syn}}
\newcommand{\Ipp}{I_\mathrm{pp}}
\newcommand{\Ibrem}{I_\mathrm{brem}}
\newcommand{\Iline}{I_\mathrm{line}}
\newcommand{\neRel}{n_e^\mathrm{rel}}
\newcommand{\npRel}{n_p^\mathrm{rel}}

\newcommand{\BtwoAve}{\overline{B_2}}
\newcommand{\BtwoAlpha}{\overline{B_2^\alpha}}

\newcommand{\pcc}{cm$^{-3}$}
\newcommand\Msun{\mathrm{M}_{\odot}}

\newcommand{\Bamp}{B_\mathrm{amp}}
\newcommand{\EffDSA}{{\cal E_\mathrm{DSA}}}

\newcommand{\Kep}{K_\mathrm{ep}}
\newcommand{\np}{n_p}
\newcommand\tSNR{t_\mathrm{SNR}}
\newcommand\dSNR{D_\mathrm{SNR}}
\newcommand\EnSN{E_\mathrm{SN}}
\newcommand\Mej{M_\mathrm{ej}}
\newcommand{\RFS}{R_\mathrm{FS}}

\newcommand{\fsk}{f_\mathrm{sk}}
\newcommand{\VelFS}{V_\mathrm{FS}}

\newcommand{\NEI}{non-equilibrium ionization}
\newcommand{\CD}{contact discontinuity}
\newcommand{\xx}[1]{\!\times\!10^{#1}}

\newcommand{\DSA}{diffusive shock acceleration}

\newcommand{\kmps}{km s$^{-1}$}
\newcommand{\NL}{nonlinear}

\newcommand{\gamrays}{$\gamma$-rays}

\newcommand{\alf}{Alfv\'en}
\newcommand{\SNRJ}{SNR RX J1713.7-3946}
\newcommand{\SNRJm}{RX J1713.7-3946}
\newcommand{\SNRJmm}{J1713}
\newcommand{\SC}{self-consistent}

\newcommand{\Rtot}{R_\mathrm{tot}}
\newcommand{\Rsub}{R_\mathrm{sub}}
\newcommand{\muG}{$\mu$G}

\newcommand{\BCSM}{B_\mathrm{0}}

\newcommand{\Epmax}{E^\mathrm{max}_p}

\newcommand{\be}{\begin{eqnarray}}
\newcommand{\ee}{\end{eqnarray}}

\newcommand{\rel}{relativistic}

\newcommand{\syn}{synchrotron}
\newcommand{\synch}{synchrotron}
\newcommand{\pion}{pion-decay}
\newcommand{\IC}{inverse-Compton}
\newcommand{\brem}{bremsstrahlung} % DCE
\newcommand{\brems}{bremsstrahlung} % DCE

\begin{document}

  \title{Efficient cosmic ray acceleration, hydrodynamics, and
  Self-consistent Thermal X-ray Emission applied to SNR RX J1713.7-3946}

\author{Donald C. Ellison,\altaffilmark{1} Daniel
  J. Patnaude,\altaffilmark{2} Patrick Slane,\altaffilmark{2} and John
  Raymond\altaffilmark{2}}

\altaffiltext{1}{Physics Department, North
Carolina State University, Box 8202, Raleigh, NC 27695, U.S.A.;
don\_ellison@ncsu.edu, avladim@ncsu.edu}

\altaffiltext{2}{Smithsonian Astrophysical Observatory, MS-3, 60 Garden
  Street, Cambridge, MA 02138, USA}

\begin{abstract}
We model the broad-band emission from \SNRJ\ including, for the first
time, a consistent calculation of thermal X-ray emission together with
non-thermal emission in a \NL\ \DSA\ (DSA) model.
Our model tracks the evolution of the SNR including the plasma
ionization state between the forward shock and the \CD. We use a plasma
emissivity code to predict the thermal X-ray emission spectrum assuming
the initially cold electrons are heated either by Coulomb collisions
with the shock heated protons (the slowest possible heating), or come
into instant equilibration with the protons.  For either electron
heating model, electrons reach $\gtrsim 10^7$\,K rapidly and the X-ray
line emission near 1 keV is more than 10 times as luminous as the
underlying thermal \brem\ continuum.  Since recent {\it Suzaku}
observations show no detectable line emission, this places strong
constraints on the unshocked ambient medium density and on the \rel\
electron to proton ratio.  For the uniform circumstellar medium (CSM)
models we consider, the low densities and high \rel\ electron to proton
ratios required to match the {\it Suzaku} X-ray observations {\it
definitively rule out \pion} as the emission process producing GeV-TeV
photons. We show that leptonic models, where \IC\ scattering against the
cosmic background radiation dominates the GeV-TeV emission, produce
better  fits to the broad-band thermal and
non-thermal observations in a uniform CSM.
\end{abstract}

\keywords{ acceleration of particles, shock waves, ISM: cosmic rays,
           ISM: supernova remnants, magnetic fields, turbulence}

\section{Introduction}
The supernova remnant \SNRJm\ (G347.3-0.5) has been detected at
photon energies ranging from radio to TeV \gamrays. The GeV-TeV
detections in particular make this SNR an important test-bed for models
of particle acceleration in astrophysical shocks, and a large number of
fits to the data have been presented with an array of environmental and
particle acceleration parameters. Invariably, parameters are found that
allow good fits to the non-thermal observations (or some sub-set of the
observations). 
A critical question for cosmic-ray (CR) origin concerns
the production of the GeV-TeV \gamrays. 
Are these \gamrays\ primarily from \IC\ (IC) emission from \rel\
electrons, or \pion\ emission from the interaction of \rel\ hadrons with
the ambient medium? 
Models with good fits to the TeV emission with either \IC\ or \pion\
have been presented
\citep[e.g.,][]{PMS2006,BV2008,Tanaka2008,MAB2009,ZA2009,YKK2009},
and strong but conflicting claims for or against one or the other
scenario, based on broad-band continuum observations, have been made
\citep[e.g.,][]{KW2008,Plaga2008,BV2009}.
We find that it is hard to discriminate on the basis of continuum
emission alone, but that thermal X-ray line emission can easily
differentiate between IC and \pion\ models because \pion\ requires a
high proton number density, $n_p$, and the thermal emission scales as
$n_p^2$.

Until now,  fits to the broad-band emission that incorporate
nonlinear diffusive shock acceleration (DSA) have not accurately
accounted for the thermal X-ray emission that might be present.  We do
this here for a SNR evolving in a uniform circumstellar medium (CSM)
with no density enhancements as might occur with a pre-SN dense shell,
nearby molecular cloud, etc.
We find that the lack of observed thermal line emission {\it eliminates
pion-decay} as the source of TeV emission in models with uniform
circumstellar media.

The essential elements of our CR-hydro-NEI model have been presented in
\citet{EC2005,EPSBG2007,PES2009}, and references therein. We couple a
one-dimensional hydrodynamic simulation of an evolving SNR with \NL\
\DSA.  The ionization structure, free electron number density, and
electron temperature in the evolving interaction region between the
forward shock (FS) and contact discontinuity (CD) are determined with a
self-consistent treatment of the nonequilibrium ionization (NEI).
We couple our computed nonequilibrium ionization fractions of heavy
elements to an updated version of the \citet{RS1977} plasma emissivity
code to compute the thermal X-ray emission.

Simultaneously, the shock accelerated, non-thermal electron and proton
spectra are calculated, evolved, and used to determine the \synch, IC,
non-thermal \brems, and \pion\ emission from the SNR. We therefore
obtain, for the first time, consistent thermal and non-thermal emission
in an evolving SNR.

\section{Model}
\label{sec:model}
Any reasonably complete broad-band model of a SNR has a host of
parameters. \SNRJ\ is no exception and in this paper we do not present a
full parameter search. Instead, we concentrate on three essential
coupled components: (1) the SNR hydrodynamics, (2) \NL\ DSA, and (3)
\NEI.

Following the majority of work on \SNRJ, we assume an age
$\tSNR\simeq1600$\,yr, and a distance $\dSNR \simeq 1$\,kpc. Using the
observed angular size, $\dSNR$ implies a forward shock radius
$\RFS\simeq 8.7$\,pc. 
While \SNRJ\ is believed to be a core-collapse SN,
we again follow the majority of work on this remnant and assume a
uniform CSM with constant proton number density, $\np$, and constant
unshocked magnetic field, $\BCSM$. We will present models where a pre-SN
wind is assumed in future work. Besides $\np$ and $\BCSM$, the
following environmental parameters are required to model the SNR
evolution:
the SN explosion energy, $\EnSN$, the ejecta mass, $\Mej$ (we
  assume an exponential mass distribution for the ejecta), and the
  temperature of the unshocked CSM, $T_0$.

We show models with two sets of parameters. In the ``hadronic'' model,
the parameters are such that the GeV-TeV emission is dominated by \pion,
while in the ``leptonic'' model, IC produces the GeV-TeV emission. In
both cases, the parameters are chosen to simultaneously match the HESS
TeV observations \citep[][]{Aharonian2007} and the {\it Suzaku}
X-ray continuum \citep[][]{Tanaka2008}.
The hadronic and leptonic names refer to the particles, protons or
electrons, mainly responsible for the GeV-TeV emission. As we show
below, both models place the majority of the accelerated particle energy
in protons, not electrons.
 
We include an amplification factor, $\Bamp$, for the shocked magnetic
field. In our simple {\it ad hoc} model of magnetic field amplification
(MFA), the compressed magnetic field immediately behind the shock is
increased by a factor $\Bamp$. The amplified downstream field is then
evolved in the downstream region as described in \citet{EPSBG2007}.  For
more \SC\ models of MFA see, for example, \citet{VEB2006};
\citet{CBAV2008}; \citet{VBE2009}.

To model the nonthermal radiation, we need additional parameters for
\NL\ DSA.\footnote{The model of nonlinear DSA we use here is based on
the semi-analytic model developed by \citet{BGV2005} and \citet{AB2006}.
In our implementation, we fix the acceleration efficiency rather
than the injection fraction as is done by Blasi and co-workers. While
this difference may have important consequences during the early stages
of the SNR evolution when the FS Mach number is extremely large
\citep[see][]{BE99}, it makes no significant difference to the integrated
spectra at the later times we show here.}
These are the acceleration efficiency, $\EffDSA$ (i.e., the
instantaneous fraction of shock ram kinetic energy flux placed in
superthermal protons), the \rel\ electron to \rel\ proton ratio,
$\Kep$,\footnote{Note that $\Kep$ sets the post-shock \rel\
electron density given the post-shock \rel\ proton density. The
post-shock thermal electron density, which determines the \brems\
continuum and the X-ray line emission, is set by the densities and
ionization states of the post-shock hydrogen and heavier elements. The
model parameters $\Kep$ and $n_p$ are independent.}
the maximum energy the protons obtain $\Epmax$, and a factor, $\cutoff$,
characterizing the shape of the turnover region around $\Epmax$. We
determine $\Epmax$ by limiting the acceleration when the acceleration
time matches the SNR age or when the upstream diffusion length matches
some fraction, $\fsk$, of the shock radius, whichever comes
first.\footnote{The diffusion length in the FS precursor is determined
assuming ``Bohm diffusion,'' where a particle's mean free path is
on the order of its gyroradius.}
The factor $\cutoff$ smoothes the particle spectrum around $\Epmax$
mimicking the effects of particle escape \citep[see, for example,
][]{ZP2008}. The above parameters are fully defined in \citet{EDB2004}
and \citet{EC2005}.
The efficiency of DSA has been directly measured at the quasi-parallel
Earth bow shock with $\EffDSA \gtrsim 0.25$ \citep[][]{EMP90}. 
Indirect evidence, based on particular models, suggests that 
the efficiency in some young SNRs, at least in some regions of the FS,
can be 50\% or more
\citep[e.g.,][]{VBK2003,WarrenEtal2005,HelderVinketal2009}.

For the thermal X-ray emission, we assume cosmic abundances and compare
two extremes for heating the initially cold electrons.  The slowest
possible heating is from Coulomb collisions and the fastest is instant
equilibration between electrons and protons, presumably produced by
wave-particle interactions. 
For shock speeds above $\sim 1000$\,\kmps, it has been suggested that
electrons are heated very rapidly to $kT \sim 0.3$\,keV by
lower hybrid waves, after which
continued heating to $kT \sim 1$\,keV proceeds through Coulomb collisions
\citep[e.g.,][]{GLR2007}. Since, as we show below, Coulomb collisions
alone rapidly heat the gas to $\sim 0.3$\,keV, any difference between
lower hybrid wave heating and Coulomb heating would only be important
for UV and optical lines, so pure Coulomb models
are appropriate for the X-ray emission.
 
It is important to note that, in our CR-hydro-NEI model for the
interaction region between the CD and FS, including X-ray line emission
only requires two additional assumptions. One is the CSM elemental
abundance and the other is the electron heating model. For Type Ia SNe,
and a wide range of low-to-moderate mass core-collapse SNe, it is
reasonable to assume solar abundances for the CSM
\citep[e.g.,][]{ChiosiMaeder86,KudPuls2000}. The two heating extremes we
consider cover all likely possibilities.

\begin{figure}
\epsscale{1.05}
\plotone{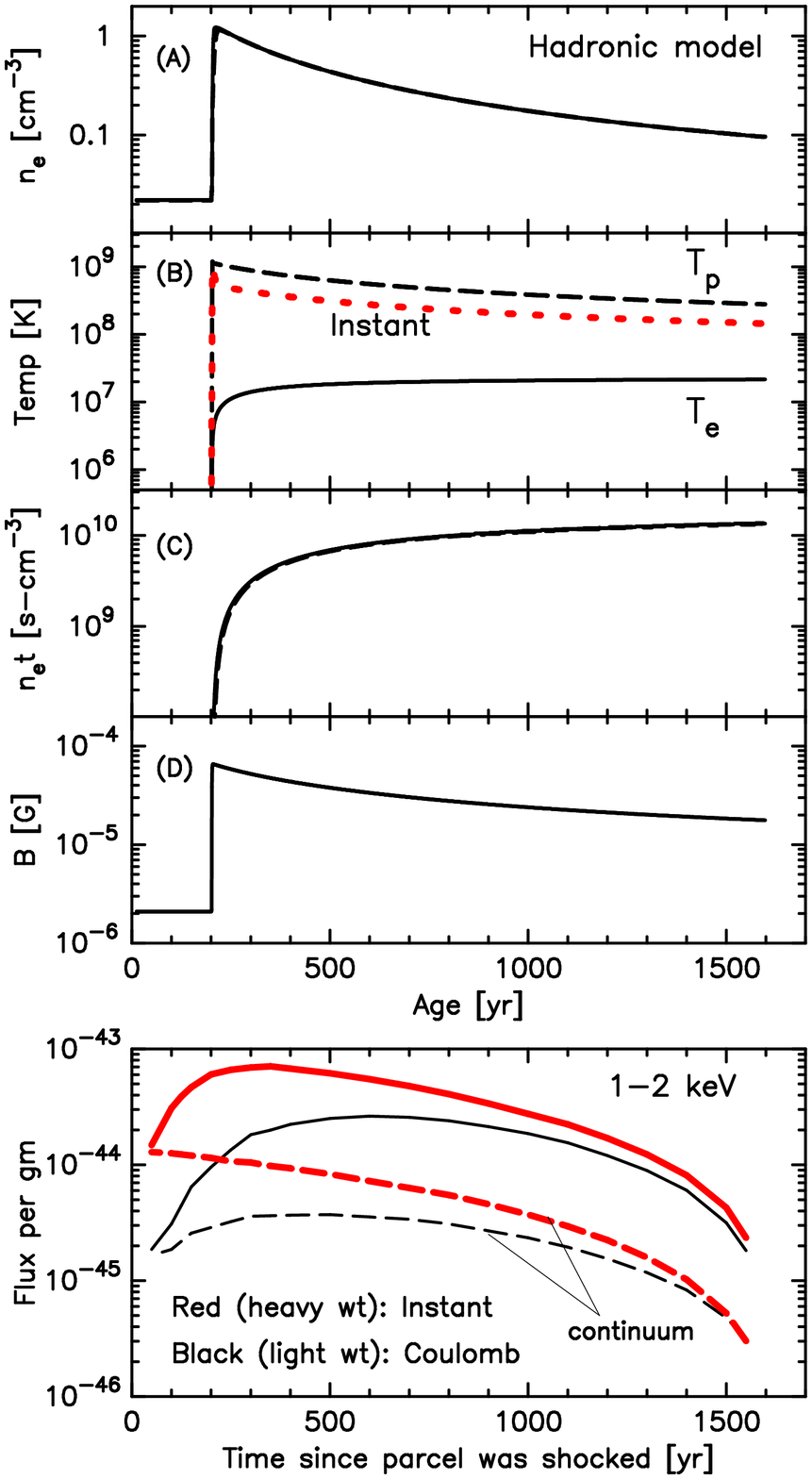}  %{f1_5stack_N.eps}          % Fig 1
\caption{ The top four panels show the free electron density, $n_e$, the
temperature, the ionization age, $n_et$, and the magnetic field in a
parcel of gas first shocked at 200\,yr.  In panels (A) and (C), the
solid curves are for Coulomb equilibration and the dashed curves (barely
visible) are for instant equilibration. In panel (B), the dashed curve
is the proton temperature and the solid curve the electron temperature
assuming Coulomb equilibration. The dotted red curve in (B) shows the
equal electron and proton temperatures assuming instant equilibration.
In the bottom panel, the solid curves show the total emitted flux, at
  $\tSNR=1600$\,yr, per arbitrary unit mass, in the band $1-2$ keV from
  parcels of gas shocked at previous times. The dashed curves in the
  bottom panel show the corresponding flux from the \brems\
  continuum. The parameters are for our hadronic model with
  $n_p=0.2$\,\pcc.
\label{fig:5stack}}
\end{figure}

\section{Results}
For our leptonic model we assume
$n_p=0.05$\,\pcc,
$B_0=3$\,\muG,
$\EffDSA=0.25$,
$\Kep=2\xx{-2}$,
$\Bamp=1$,
$\fsk=0.1$, and
$\cutoff=1$.
For the hadronic model,
$n_p=0.2$\,\pcc,
$B_0=2$\,\muG,
$\EffDSA=0.5$,
$\Kep=7\xx{-4}$,
$\Bamp=5$,
$\fsk=0.05$, and
$\cutoff=1$.  
In both models, the values for $\EnSN$ and $\Mej$ are varied with $n_p$
to obtain $\RFS\sim 8-10$\,pc at $\tSNR=1600$\,yr. Thus, for the leptonic
model,
$\EnSN=1\xx{51}$\,erg and
$\Mej=3\,\Msun$, 
while the hadronic model uses
$\EnSN=2\xx{51}$\,erg and
$\Mej=1.4\,\Msun$.\footnote{The value $\Mej=1.4\,\Msun$ is not meant to
  imply that we believe \SNRJ\ originated from a Type Ia supernova.}
For a particular $n_p$, other combinations of $\EnSN$ and $\Mej$ giving
$\RFS\simeq 8-10$\,pc at 1600\,yr yield similar results.
In all cases, we assume $T_0=10^4$\,K.\footnote{As long as
$T_0\lesssim 10^6$\,K, the unshocked temperature only weakly
influences our results.} 

At the end of the simulation, we obtain for the leptonic (hadronic)
model:
the forward shock radius $\RFS \simeq 9.3 \ (8.8)$\,pc;
the forward shock speed $\VelFS\simeq 3000 \ (2300)$\,\kmps;
the magnetic field immediately behind the FS $B_2\simeq 10 \ (36)$\,\muG;
the overall FS compression ratio $\Rtot \simeq 4.6\ (5.6)$,
the subshock compression ratio $\Rsub \simeq 3.98 (3.86)$,
the fraction of SN explosion energy placed in CR ions $\simeq  0.13 \
(0.4)$, and
the mass swept up by the FS $ \simeq  6 (19)\,\Msun$.

In Fig.~\ref{fig:5stack} we illustrate the properties of our
CR-hydro-NEI model by following particular parcels of plasma. 
In the top four panels we show, for our hadronic model, the free electron
number density, $n_e$, the electron and proton temperatures, the
ionization parameter or age, $n_e t$ ($t$ is the time since the parcel
was shocked), and the magnetic field in a parcel of plasma that is
overtaken by the FS at 200\,yr. 
The red
dotted curve in panel (B) gives the temperature assuming instant
equilibration.
Even though $T_e/T_p \lesssim 0.1$ throughout the simulation for Coulomb
equilibration, $T_e$ approaches $10^7$\,K ($\sim 850$\,eV)
rapidly before leveling out.

In the bottom panel of Figure~\ref{fig:5stack}, we plot the thermal
X-ray emission between 1 and 2 keV, for both instant and Coulomb
equilibration, at the end of the simulation for parcels of plasma
shocked at previous times. The dashed curves are the continuum emission
between 1 and 2 keV and the total emission (solid curves), including
lines, stands well above this regardless of the electron equilibration.
As the left end of the bottom panel shows, at $\tSNR \simeq 1600$\,yr,
plasma that was shocked $\gtrsim 200$ years earlier is sufficiently
ionized to produce a substantial flux in lines regardless of the
electron equilibration.

\begin{figure}
\epsscale{1.05}
\plotone{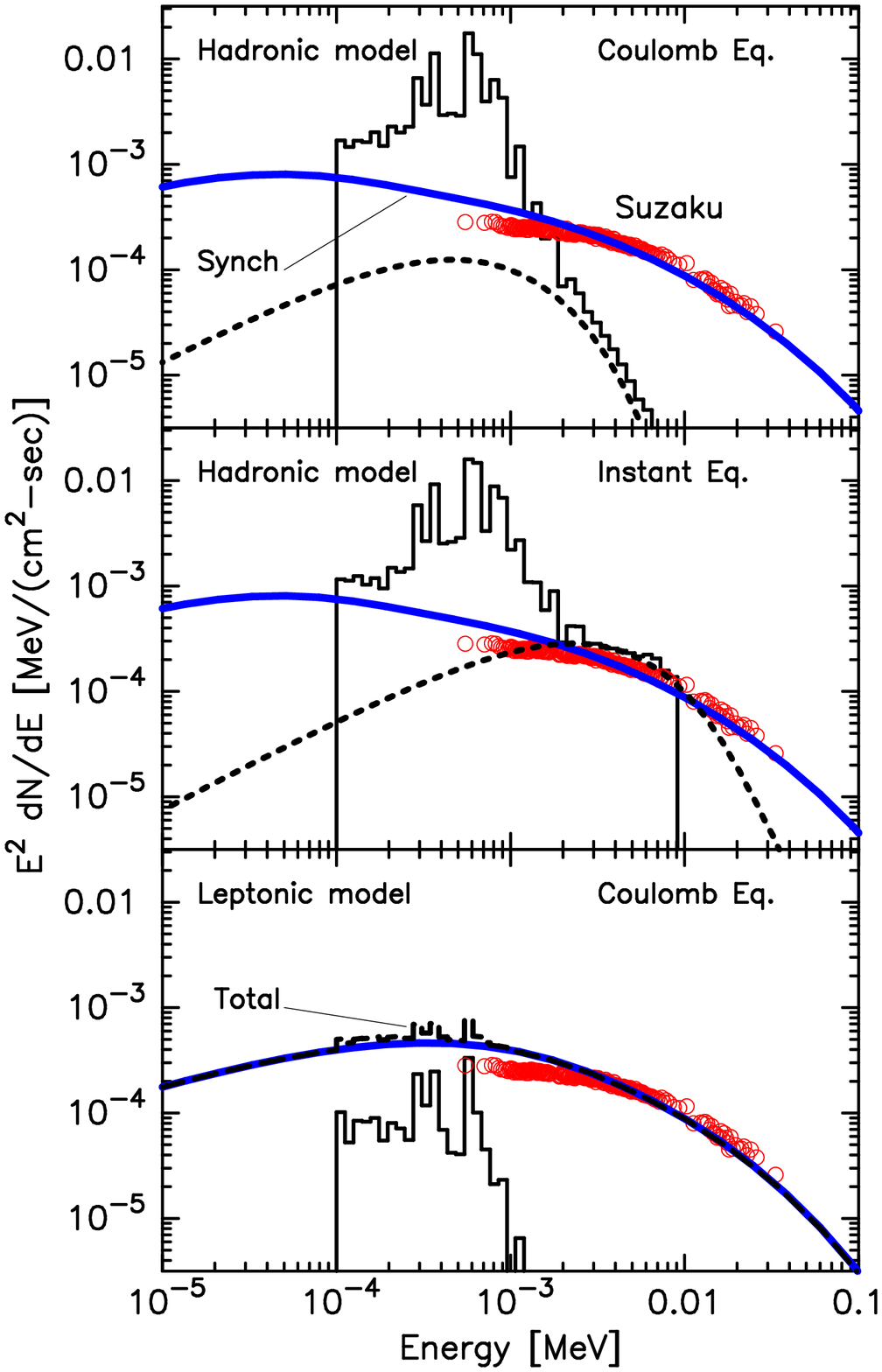}  %{f2_xray_Hadronic_Leptonic.eps}  % Fig 2
\caption{
The top two panels show fits to the {\it Suzaku} \SNRJm\ observations
  with our hadronic model for both Coulomb and instant temperature
  equilibration but ignoring the X-line emission.  The blue (heavy wt.)
solid curve
  is the \synch\ continuum, the black solid curve is the thermal
  emission (only lines above $10^{-4}$\,MeV are included), and the
  dotted curve is the underlying \brems\ continuum. The observed
  emission would be the sum (not shown in the top two panels) of the
  solid black and blue curves.  The bottom panel shows the leptonic
  model (with Coulomb equilibration) where parameters have been chosen to
  be consistent with the {\it Suzaku} observations.
For the hadronic model, the radiation intensity is multiplied by 0.95 to
match the observations. For the leptonic model, a normalization factor
of 0.2 is required to match the observations. We note that the {\it
Suzaku} data have been adjusted for interstellar extinction so no
extinction is applied to the model in this plot.
\label{fig:xray}}
\end{figure}

In the top two panels of Figure~\ref{fig:xray} we compare our hadronic
model to {\it Suzaku} observations of \SNRJmm\ \citep[][]{Tanaka2008}
for Coulomb (top panel) and instant equilibration (middle panel). The
{\it Suzaku} observations have been adjusted for interstellar extinction
and all model parameters are the same as in Fig.~\ref{fig:5stack}.
For our hadronic model, we have chosen parameters that result in \pion\
dominating the GeV-TeV emission, i.e., $n_p$ must be above some limit
and $\Kep$ must be below some limit for this to be the case.
Figure~\ref{fig:xray} makes it clear, however, that the X-ray line
emission is much stronger in the hadronic model than can be accommodated
by observations.  This is true for Coulomb equilibration even though the
\brems\ continuum remains well below the {\it Suzaku} observations.
The only way to lower this emission relative to the
synchrotron continuum would be to increase $\Kep$ or to decrease $n_p$
to values that would then no longer reproduce the observed gamma-ray
emission.
This is true regardless of the electron equilibration.  We note that
lowering $n_p$ in uniform CSM models requires lowering $\EnSN$ to
maintain $\RFS\sim 8-10$\,pc. 

We are unable to find any set of parameters that gives \pion\ dominating
the TeV emission without producing emission lines around 1 keV that are
inconsistent with the {\it Suzaku} observations.

In the bottom panel of Figure~\ref{fig:xray} we show our leptonic model
where we have chosen parameters to be consistent with the smooth {\it
Suzaku} observations.  
In addition to the parameters discussed
already, we have arbitrarily adjusted the overall normalization of
both models to match the observations. The hadronic model has been
multiplied by 0.95 and the leptonic model by 0.2. Normalization values 
$< 1$ might correspond, observationally, to a partially complete shell
morphology for the SNR, or possibly some reduction in the DSA injection
and/or acceleration efficiency over some fraction of the SNR surface
\citep[e.g.,][]{BV2008}.

\begin{figure}
\epsscale{1.05}
\plotone{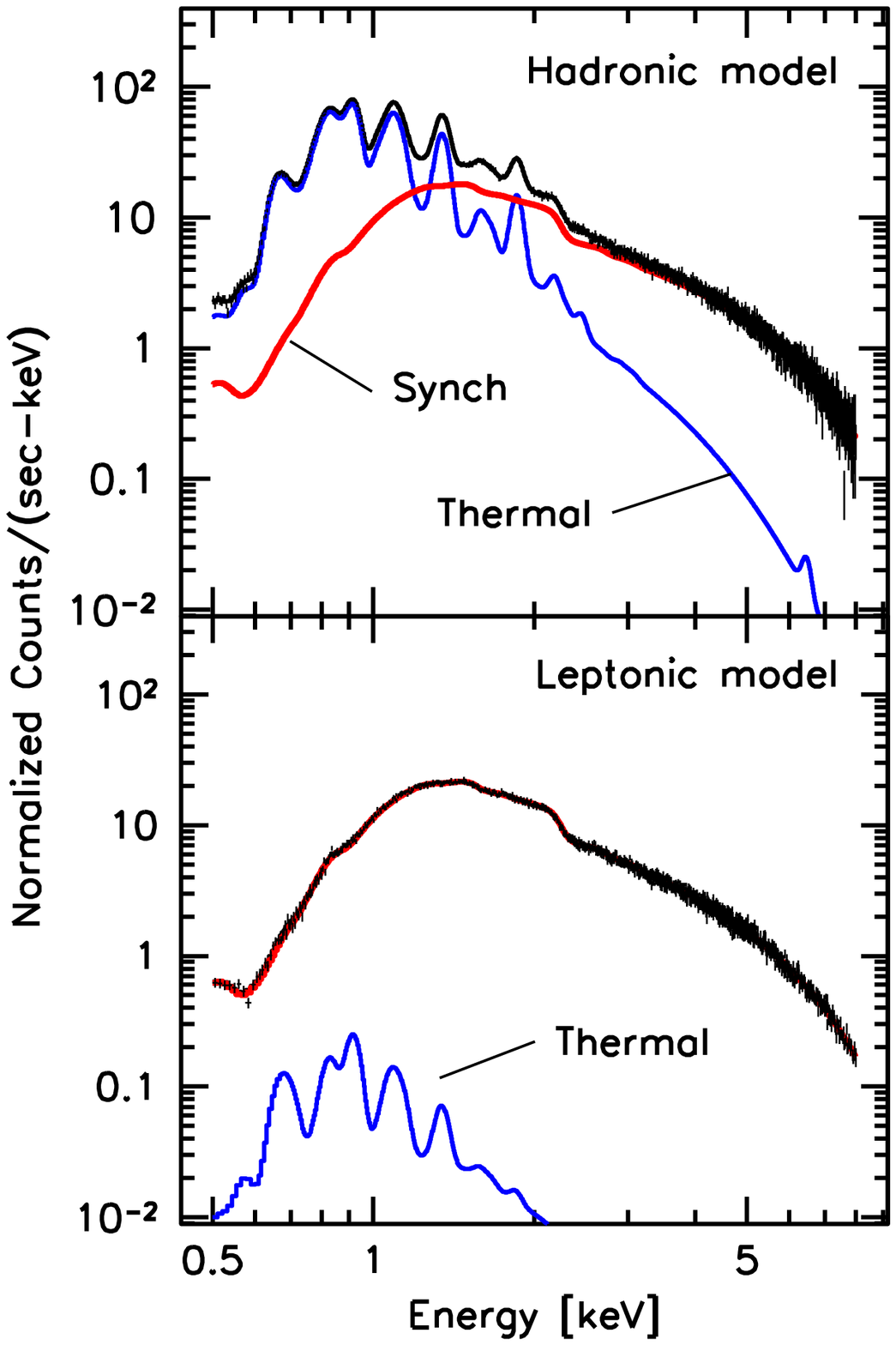}  %{Suzaku_XIS.eps} % Fig 3
\caption{Simulated {\it Suzaku} XIS spectra of RX J1713.3-3946. In the
top panel, the best fit hadronic model is shown, with $n_p = 0.2$\,\pcc,
while in the bottom panel, the best fit leptonic model is
shown, with $n_p = 0.05$\,\pcc. In both panels, the blue curve is the
contribution from the thermal X-ray emission, while the red curve is the
contribution from synchrotron emission. The spectra correspond to a
simulated 20 ks observation and are normalized to match the unabsorbed
1.0 - 10.0 keV flux of 7.65 $\times 10^{-10}$ erg cm$^{-2}$ s$^{-1}$
found by Tanaka et al.~(2009). In these simulated observations, we
assume a Galactic n$_{\mathrm H} = 7.9\xx{21}$\,cm$^{-2}$.
\label{fig:sim_Suzaku}}
\end{figure}

In Figure~3, we show our best fit hadronic and leptonic models, folded
through the {\it Suzaku} XIS instrument response.\footnote{Response
matrices are available at
http://heasarc.nasa.gov/docs/suzaku/prop\_tools/xis\_mat.html.}
For both models, we simulated 20\,ks observations of the entire SNR with
no background subtraction, assuming a Galactic column density
n$_{\mathrm{H}}$ = 7.9 $\times$ 10$^{21}$ cm$^{-2}$.  When compared to
the {\it Suzaku} observations \citep[cf., Figs.~10 or 11
in][]{Tanaka2008}, it is clear that {\it Suzaku} would have detected
lines as strong as those produced in our hadronic model had they been
present.

\begin{figure}
\epsscale{1.05}
\plotone{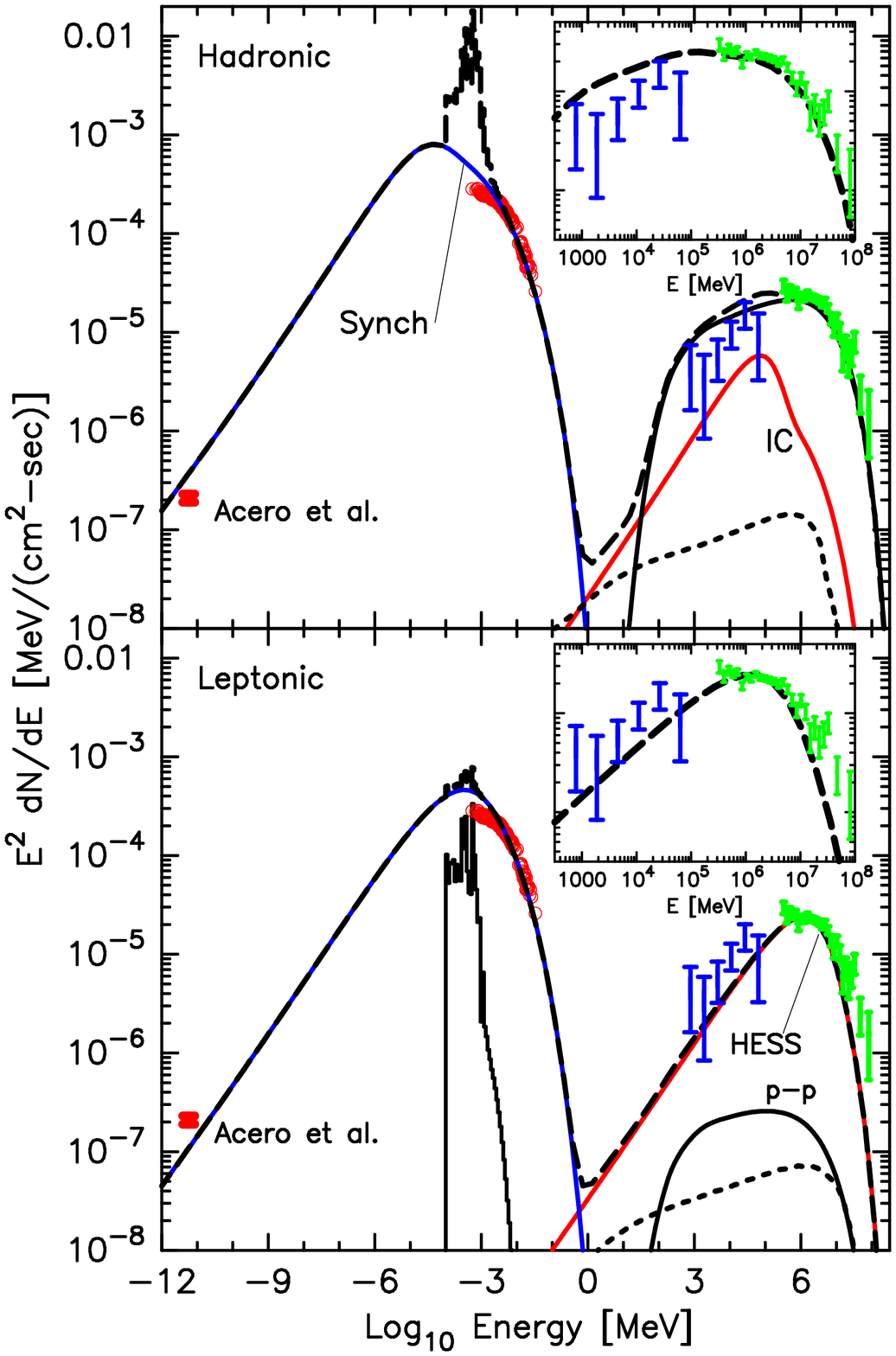}  %{f4_4frame_newHESS_LAT.eps} %f4_4frame_BB_LAT.eps}  % Fig 4
\caption{Broad-band fits to radio \citep[][]{Acero2009}, {\it Suzaku}
\citep[][]{Tanaka2008}, preliminary {\it Fermi-LAT}
\citep[][]{FunkJ17132009}, and HESS observations
\citep[][]{Aharonian2007} of \SNRJm. The top panel is our 
hadronic model and the bottom panel is our 
leptonic model. In both cases, the blue curve
is \synch, the black is \pion, the red is IC, and the dotted is
non-thermal \brems. The dashed black curve is the sum including the
X-ray line emission.  As in Fig.~\ref{fig:xray}, a normalization factor
of 0.95 (0.2) has been applied to the hadronic (leptonic) model.
\label{fig:broadband}}
\end{figure}

\begin{figure}
\epsscale{1.05}
\plotone{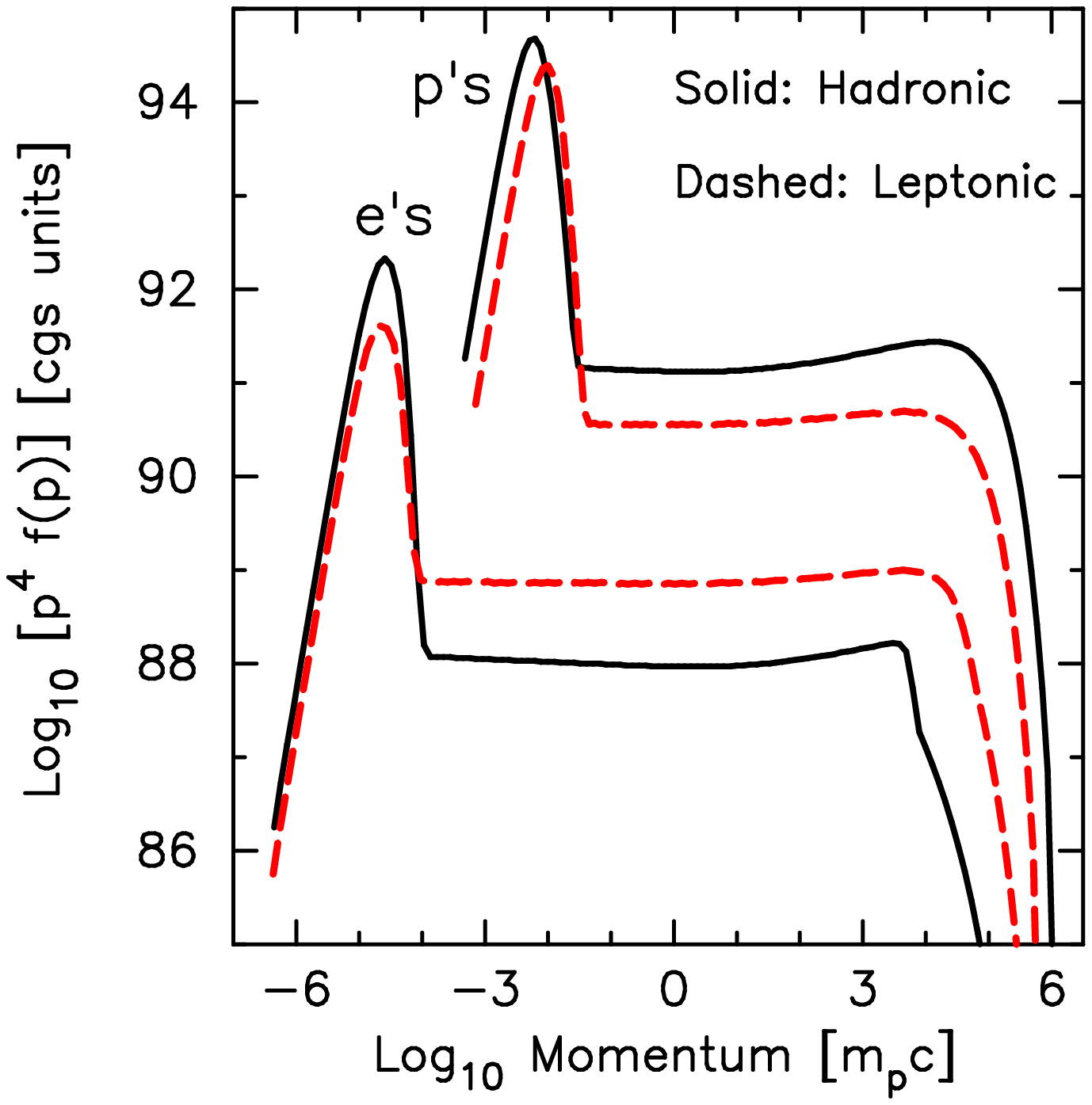}  %{fp_Had_Lep.eps}  % Fig 5
\caption{Phase-space distribution functions multiplied by $p^4$ for our
  hadronic (black solid curves) and leptonic (red dashed curves)
  models. These are integrated spectra at the end of the simulation in
  cgs units and represent the total material swept-up by the FS. 
\label{fig:fp}}
\end{figure}

In Fig.~\ref{fig:broadband}, we show broad-band fits to radio, {\it
Suzaku}, preliminary {\it Fermi-LAT}, and HESS observations of \SNRJm.
The hadronic and leptonic models both produce reasonable fits if the
thermal X-ray line emission is ignored. When the thermal X-rays are
considered, the hadronic model is excluded.
Only the cosmic microwave background is used to determine the IC
emission.

It is important to note in considering Figs.~\ref{fig:xray} and
\ref{fig:broadband} that equally good fits to the {\it continuum}
observations can be obtained with different parameter
combinations. This, and the fact that the various models that have been
applied to \SNRJm\ differ in details, accounts for the relatively small
differences in parameters we obtain compared to those obtained by other
modelers \citep[e.g.,][]{BV2008,MAB2009}. However, consistency with the
thermal X-ray line emission forces the CSM density down and $\Kep$ up so
no set of parameters can be found that result in \pion\ dominating the
GeV-TeV emission.

Characteristically of efficient DSA, the CR-hydro-NEI model produces an
overall shock compression, $\Rtot > 4$, and a subshock compression,
$\Rsub <4$.  Nevertheless, even with 50\% efficiency ($\EffDSA =0.5$),
$\Rsub$ remains large enough for electrons temperatures to be high
enough for strong line production.

The only factor we see that could lower the thermal emission
substantially in a uniform CSM model, is the abundance. If the CSM is
nearly devoid of heavy elements, thermal line emission will be
suppressed.
Depletion onto dust will cut down C, Mg, Si and Fe, but it will not
affect the O lines, which are the brightest in the model, or N or Ne.
Furthermore, a substantial fraction of the dust is destroyed once $n_e
t$ becomes a few times $10^{10}$\,s-cm$^{-3}$, so some of the refractory
elements would be liberated \citep[e.g.,][]{Williams2006}.  One does not
expect really severe depletion in the low density uniform medium, but
there could be significant dust in a red giant wind.

It is also possible that the progenitor was a Wolf-Rayet (WR) star,
and this could give anomalous abundances
\citep[e.g.,][]{Crowther2007}.
Conversion of H to He reduces the number of electrons, weakening the
line emission per unit mass by as much as a factor of two.  However, WC
and WO stars show much larger overabundances of O and Ne, which produce
the strongest lines in the spectra, so the lines would be strongly
enhanced.  In WN stars, carbon and oxygen have been converted to
nitrogen.  The O lines would be weakened, and the N VII line at 500 eV
would be luminous but badly attenuated.  The Ne IX and X lines at 922
and 1022 eV would then be the strongest in the spectrum at 0.5 to 1
times the strengths predicted.
Thus, even in the case of a progenitor wind with anomalous abundances,
we would still expect to see strong line emission in the swept-up CSM,
and this would be present in the Suzaku observations.

\section{Discussion and Conclusions}
While several authors have proposed that emission lines could be
undetectable in \SNRJmm\ because of low shock temperatures or
time-dependent ionization \citep[e.g.,][]{DAMG2009,MAB2009,BV2009}, we
find that a SNR with properties typically ascribed to \SNRJmm, expanding
in a uniform CSM with solar elemental abundances, will produce strong
X-ray emission lines when electron equilibration and \NEI\ are taken into
account.
This places constraints on the CSM density, $n_p$, and on the \rel\
electron to proton ratio, $\Kep$, to be consistent with {\it Suzaku}
observations which show a smooth X-ray \synch\ continuum with no lines.

While particular values of $n_p$ and $\Kep$ will depend somewhat on
details of various DSA and SNR models, in any uniform CSM
model the CSM must have a relatively low density and the electron to
proton ratio of shock accelerated particles must be relatively high in
order to produce a satisfactory fit to the {\it Suzaku} data.
Models where \pion\ produces the observed TeV emission require densities
that are too high and values of $\Kep$ that are too low to be consistent
with the {\it Suzaku} observations.
We note that we have actually only computed a lower limit to the line
emission since we have not included line emission from the ejecta
material heated by the reverse shock.  
If emission from a RS had been included, our conclusion that \pion\ is
excluded could only be strengthened.

Apart from minor differences, our fit to the broad-band spectrum (bottom
panel Fig.~\ref{fig:broadband}) is consistent with others
\citep[e.g.,][]{PMS2006,MAB2009} where IC dominated the TeV emission.
Our results differ substantially from the conjecture made by
  \citet{DAMG2009} that the post-shock temperature can be reduced below
  X-ray emitting temperatures in strong shocks. The conclusions of
  \citet{DAMG2009} are based on scaling arguments in the limit of
  extremely high sonic and \alf\ Mach numbers where the acceleration
  efficiency approaches 100\%. In this case, the subshock may become
  weak enough to limit heating to the values \citet{DAMG2009}
  suggest. However, Mach numbers as high as assumed in the
  \citet{DAMG2009} scalings are not obtained for reasonable ambient
  magnetic fields and other parameters normally assigned to \SNRJm. When
  nonlinear effects are fully taken into account for \SNRJmm\ parameters
  \citep[see also,][]{MAB2009}, and for acceleration efficiencies even
  modestly below 100\%, the post-shock plasma (i.e., the proton
  component) is heated more strongly than \citet{DAMG2009} suggest.

We further emphasize that there is little freedom to reduce the
thermal emission since we have calculated the NEI for the two electron
heating extremes: Coulomb collisions and instant equilibration. For both
extremes, and all cases in between, the shock heated plasma produces
strong line emission. As Fig.~\ref{fig:5stack} shows, it is not
necessary for electrons to equilibrate with protons to become hot enough
for line emission. For Coulomb collisions in our hadronic model,
$T_e/T_p$ remains less than about 0.1 for $\gg 1600$\,yr.

Once it becomes clear that X-ray emission lines will be produced
efficiently with a luminosity approximately $\propto n_p^2$, the
intensities, $I$, of all the emission processes can be roughly scaled
with the important parameters, $n_p$, $\Kep$, and the  average
downstream field $\BtwoAve$ as:
\begin{equation}
\Iic\propto \neRel \propto \npRel \Kep
\ ;
\end{equation}
\begin{equation}
\Isyn\propto \neRel \BtwoAlpha \propto \npRel \Kep 
\BtwoAlpha
\ ;
\end{equation}
\begin{equation}
\Iline \propto \Ibrem\propto n_p^2
\ ;
\end{equation}
and
\begin{equation}
\Ipp\propto \npRel n_p \propto \Facc n_p^2
\ .
\end{equation}
Here, the superscript ``rel'' indicates the number density of \rel\
particles capable of producing the observed radiation. The factor
$\Facc$ is some function of the DSA efficiency, i.e., the fraction of
ambient protons turned into \rel\ protons capable of producing GeV-TeV
emission ($\npRel \propto \Facc n_p$). We also assume that the \rel\
protons producing \pion\ are drawn from the same population as the
target protons.
The expression for $\Isyn$ assumes the underlying electron spectrum is a
power law, $dN/dE \propto E^{-\sigma}$, with $\sigma=2\alpha -1$.

If the TeV emission is from \pion, then the ratio $\Isyn/\Ipp$ is fixed
by the observations and
\begin{equation}
\Isyn/\Ipp \propto \Kep \BtwoAlpha / n_p \equiv G
\ ,
\label{eq:GG}
\end{equation}
where $G$ is some constant determined by either the radio or X-ray
\syn\ observations. If $G$ is set by radio observations, radiation
losses don't play a role.
To hide the X-ray lines, we need to increase 
\begin{equation}
\Isyn/\Iline \propto \npRel \Kep \overline{B_2^\alpha}/ n_p^2 =
\npRel G/ n_p \propto \Facc G
\ .
\label{eq:synRatio}
\end{equation}
Thus, the only parameter that can change the relative intensity
$\Isyn/\Iline$  is the DSA efficiency.
The X-ray line to synchrotron continuum ratio can be changed by
  changing the magnetic field, but the absolute ratio of X-ray lines to
  gamma rays is basically fixed in the hadronic scenario.  From
  equations 3 and 4,
\begin{equation}
\Iline/\Ipp  \propto  1/\Facc
\ ,
\label{eq:linepp}
\end{equation}
and for the hadronic model (top panel in Fig.~\ref{fig:broadband}),
$\Iline/\Ipp$ is more than an order of magnitude too large compared to
observations to be accommodated.  Changing $\Facc$ and/or $B$ cannot
hide the lines if $\np$ is too large.

Of course, the situation is more complicated for several reasons. (1)
The line emission depends importantly on the SNR evolution (i.e., the
ionization age; Fig.~\ref{fig:5stack}) and the CSM composition. (2) The
factor $\Facc$ depends on the shock dynamics, the magnetic field, and
uncertain details of NL DSA. Furthermore, since radiation losses are
important for \rel\ electrons but not \rel\ protons, $\neRel/\npRel \neq
\Kep$ at high energies and $\Kep$ depends on $B$ for X-ray \syn\
emission. This will change the $\Isyn/\Iline$ scaling at X-ray energies
but will not change the relative intensities of radio vs. \pion\
emission.
(3) The detailed fits to the {\it Suzaku}, {\it Fermi-LAT}, and HESS
data depend critically on the {\it shape} of the underlying electron and
proton spectra in the turnover region, and on the SNR magnetic field
morphology. Despite these complications, Eqs.~(\ref{eq:synRatio}) and
(\ref{eq:linepp}), must largely control the overall scaling.

For our hadronic model shown in Figs.~\ref{fig:xray} and
\ref{fig:broadband}, we have chosen particular values of $n_p$, $\Kep$,
and $\Bamp$ to match the shape and relative normalization of the radio,
X-ray, and TeV observations. For the acceleration efficiency, we have
used $\EffDSA=0.5$, i.e., 50\% of the instantaneous FS ram kinetic
energy flux is put into \rel\ protons. While there is little indication
that larger values of $\EffDSA$ occur in SNRs, we have explored $\EffDSA
> 0.7$ and find a poorer match to the broadband observations and no
improvement in the hadronic fit to the X-ray lines.
One reason for this is that, in \NL\ DSA, an increase in acceleration
efficiency must be accompanied by an increase in the overall shock
compression ratio, $\Rtot$. This translates to an increase in the
downstream plasma density, a decrease in the electron temperature
equilibration time, and stronger X-ray line production. Furthermore,
increasing the acceleration efficiency also increases $\BtwoAve$ due to
compression and possibly more by MFA. Because of changes in $\BtwoAve$,
increases in $\EffDSA$ are constrained by equation~\ref{eq:GG}.

On the other hand, it is easy to show that lowering $\EffDSA$ below 0.1
is also inconsistent with the broad-band observations.

We emphasize again that the modeled shape of the high-energy
turnover is both critical and uncertain.  For IC and \synch, the shape
depends on the competition between acceleration and radiation loss
timescales in the acceleration region.  The turnover will be further
modified by radiation losses as the electrons evolve behind the shock
and by diffusion of the high-energy electrons into regions of different
density and magnetic field.  In fact, high-energy electrons might
diffuse away from regions of high magnetic field, reducing their \synch\
emission while they still emit IC,
For \pion, the turnover depends on the maximum
energy the FS can produce which depends on the self-generated diffusion
of the highest energy, escaping particles.

Since these effects are yet to be described precisely, all existing SNR
models, including ours, make arbitrary approximations that importantly
influence the turnover shape. The fit to the shape of the {\it
Fermi-LAT} and HESS observations is determined largely by $\fsk$,
$\cutoff$, and $\Bamp$. The detailed fit to the shape of the {\it
Suzaku} observations depends largely on $\cutoff$ and $\Bamp$.

Other effects may be important as well.  The \synch\ spectrum might be
hardened in the turnover region by stochastic effects, as described in
\citet{BykovDots2008}. Furthermore, as suggested by several authors
\citep[e.g.,][]{PMS2006}, a photon source in addition to the cosmic
microwave background might improve  the IC fit to the
highest energy HESS points. 

In contrast to the shapes of the radiation spectra near their
maximum energies, the relative normalizations of \synch, IC, \pion, and
thermal X-ray emission are less uncertain because they depend more
concretely on basic parameters. 
We believe that none of the
approximations in our CR-hydro-NEI model are significant enough to
change our basic conclusion: the constraints on ambient density and
$\Kep$ from thermal X-ray emission rule out \pion\ as the mechanism
producing TeV emission in  models with a uniform CSM.

The fact that electrons are likely producing the highest energy photons
 observed from \SNRJm, does not lead us to suggest that protons are
 absent or less energetic. Our leptonic model assumes that at any
 instant 25\% of the shock ram kinetic energy flux goes into \rel\
 protons while less than 1\% goes into \rel\ electrons. 
Electrons are observed simply because they radiate more
 efficiently than protons in low density media.
As Fig.~\ref{fig:fp} shows, the maximum proton energy is similar in our
 two models, i.e., $\Epmax \simeq 10^{14}$\, eV. The increase in $\Epmax$
 from the higher shocked magnetic field in the hadronic model (e.g.,
 $B_2 \simeq 36$\,\muG\ at the end of the simulation vs. $B_2 \simeq
 10$\,\muG\ for the leptonic model) is partially offset by the smaller
 $\fsk$ factor ($\fsk = 0.05$ for the hadronic model while $\fsk = 0.1$
 for the leptonic model).
The electron maximum energy is about a factor of 10 higher in the
leptonic model, and the shapes of the electron spectra are different,
due to the effects of radiation losses.

One result of our leptonic model, which integrates emission over
the entire remnant, that may conflict with observations is the low
shocked magnetic field. A low $B_2$ favors the leptonic model and we
obtain $B_2\simeq 10$\,\muG\ for the parameters used here.
Much higher estimates for $B_2$ have been obtained for thin X-ray
  filaments where the sharp X-ray edges and/or rapid time variations are
  attributed to strong \syn\ losses \citep[e.g.,][]{Uchiyama07}. Our
  uniform CSM assumption cannot descibe filaments and it is possible
  that more complicated, multi-component models could account for
this.
For example, if the \syn\ emission originates from a smaller region than
the IC emission (due, for example, to a strong but compact postshock
magnetic field), then a larger field strength would be possible for a
given inverse-Compton flux.

We have been careful to emphasize that we only consider a uniform CSM in
this paper. While \pion\ is eliminated in this simplest case, \SNRJ\ is
certainly more complex. As {\it Fermi-LAT} observations improve with
time, the shape of the combined {\it Fermi-LAT} and HESS observations
may indicate that the GeV-TeV emission is, in fact, hadronic in
origin. This will require some multi-component model where \rel\ protons
interact with a high density target but care must still be taken to
avoid inconsistency with the {\it Suzaku} observations. If the FS runs
into a high density shell, strong X-ray lines are likely to be produced
along with the enhanced \pion\ emission. 
If the highest energy protons escape upstream from the FS and impact a
high density medium before the material is shock heated, \pion\ emission
may be strong without strong accompanying X-ray line emission
\citep[e.g.,][]{LKE2008}.

\acknowledgments We thank Matthew Baring, Roger Blandford, Andrei Bykov,
Stefan Funk, Matthieu Renaud and Andrey Vladimirov for helpful
discussions on this work. We thank Yasunobu Uchiyama for furnishing the
{\it Suzaku} data and Stefan Funk for furnishing the {\it Fermi-LAT}
and HESS data. D.C.E acknowledges support from NASA grants
ATP02-0042-0006, NNH04Zss001N-LTSA, and 06-ATP06-21. P.O.S. and
D.J.P. acknowledge support from NASA Contract NAS8-03060.  P.O.S. and
D.C.E. acknowledge support from NASA Grant NNX09AT68G.  The authors are
grateful to the KITP in Santa Barbara where part of this work was done
when the authors were participating in a KITP program.  D.C.E. also
thanks KIPAC and D.J.P. acknowledges travel support from a SI Endowment
Grant.

\clearpage

\end{document}